\documentclass[aps,pra,twocolumn,showpacs,floatfix,superscriptaddress]{revtex4-1}
\usepackage{color}
\usepackage{amsmath,amssymb,amsfonts,bbm,graphicx,hyperref,times,url}
\def\Tr{\hbox{Tr}}
\def\sigmaCM{\boldsymbol{\sigma}}

\begin{document}
\title{Unravelling the noise: the discrimination of wave function collapse models under time-continuous measurements}
%\title{Optimal quantum estimation of fundamental decoherence induced by wavefunction collapse in levitated optomechanics}
\author{Marco G. Genoni}
\affiliation{Quantum Technology Lab, Dipartimento di Fisica, Universit\'a degli Studi di Milano, 20133 Milano, Italy}
\affiliation{Department of Physics and Astronomy, University College London, Gower Street, London WC1E 6BT, UK}
\author{O. S. Duarte}
\affiliation{Department of Physics and Astronomy, University College London, Gower Street, London WC1E 6BT, UK}
\affiliation{S\~{a}o Carlos Institute of Physics, University of S\~{a}o Paulo, PO Box 369, 13560-970 S\~{a}o Carlos, SP, Brazil}
\author{Alessio Serafini}
\affiliation{Department of Physics and Astronomy, University College London, Gower Street, London WC1E 6BT, UK}

\begin{abstract}
Inspired by the notion that environmental noise is in principle observable, whilst fundamental noise due to 
spontaneous localisation would not be, we study the estimation of the diffusion parameter 
induced by wave function collapse models under continuous monitoring of the environment.
We take into account finite measurement efficiencies and, 
in order to quantify the advantage granted by monitoring, we analyse the quantum Fisher information 
associated with such a diffusion parameter, identify optimal measurements in limiting cases, and 
assess the performance of such measurements in more realistic conditions.
\end{abstract}

\maketitle
\section{Introduction}
%\noindent {\em Introduction.} - 
Spontaneous localization models \cite{Bassone}, in their many flavours and variations, 
were introduced from the late eighties primarily as an attempt to unify the dynamics of microscopic and macroscopic systems, 
encompassing measurement apparata, which customary quantum mechanics only describes through ad hoc prescriptions
that cannot be relied to the fundamental dynamical principles. 
While such models reproduce quantum and classical mechanics in the extreme regimes of few ($\lesssim10^{6}$) and very many 
($\gtrsim10^{18}$) elementary constituents, they do deviate substantially from standard quantum mechanics in the intermediate mesoscopic regime. 
As molecular interferometry \cite{Gerlich,Arndt} and quantum opto-mechanics, 
especially in the levitating paradigm \cite{Chang,Oriol2010,Gieseler}, 
are swiftly advancing into this mesoscopic middle ground, there is currently a lively interest in designing 
and carrying out experiments that would 
falsify either standard quantum mechanics or its spontaneously 
localized variants \cite{Oriol2011,Bahrami14,vinante,nimmrichter,Li15,Goldwater15,Abdi16}.
 
In a nutshell, spontaneous localization models postulate the presence of an additional stochastic term in the Schr\"odinger equation, that would be responsible for the wave-function collapse and the perceived discontinuous dynamics of 
quantum projective measurements. This would essentially imply the existence of a source of ``fundamental'' decoherence, in the 
form of momentum dissipation, acting on a mesoscopic system, such as a levitating opto-mechanical nanosphere. 
It has hence been recently noted that, if the sources of ``environmental'' decoherence -- due to the interaction and entanglement with the environment -- 
are well known, the additional fundamental decoherence could be directly observed by tracking 
the system's dynamics \cite{vinante,Li15,Goldwater15}.

The detection of fundamental effects over the background of environmental ones is however obviously difficult, as the two may take the same form 
and imply qualitatively similar effects. 
The primary intent of this work is emphasising that a possible distinction between fundamental and environmental decoherence is that, 
while the former is unavoidable and beyond repair, the latter can in principle be reversed through measurements: if the physical degrees of 
freedom of the environment are completely or partially accessible, one can perform measurements on them that partly restore information about the quantum state \cite{burgnote}. 
Drawing from this notion, we will hence consider the estimation of the free parameter of QMUPL (``quantum mechanics with universal position localization'') 
or of an equivalent function of the two parameters of the CSL (``continuous spontaneous localization'') 
wavefunction collapse model, under time-continuous measurements on the environment of a quantum degree of freedom, 
such as the centre of mass of a levitated nanosphere \cite{TaylorNPhoto}. The latter will be our system of reference, 
bearing in mind that similar results would apply to more general settings.
The monitoring we consider, aided by Markovian linear feedback, has the added bonus of stabilising the dynamics \cite{Levante}, so that we will be in a position to base our investigation entirely on steady state properties and not on the features of the transient dynamics, which may be more elusive to record in practice.

As a further element of novelty, we will not just consider specific empirical signatures of the different values of the collapse parameter but instead 
address their systematic, ultimate discrimination by applying quantum estimation techniques and deriving the quantum Fisher information (QFI) associated with such a parameter \cite{MatteoIJQI}. 
Thus, we will quantify exactly the advantage provided by continuous monitoring 
as a decrease in the achievable uncertainty on the parameter estimation, and hence on the discrimination between different theories. \\
The manuscript is organized as follows: in Sec. \ref{s:QET} we will introduce the basic of local quantum estimation theory, along with the formulas for classical and quantum Fisher information. In Sec. \ref{s:dynamics} we will discuss the quantum dynamics of our system, in particular presenting the stochastic master equation that describes the time-continuous monitoring of the mechanical oscillator. In Sec. \ref{s:results} we show our results on the estimation for the fundamental diffusion parameter due to to spontaneous collapses: we shall derive analytical expressions for both the QFI and the optimal final measurement for parameter discrimination in limiting instances, and show the latter performs remarkably well 
in realistic situations too. Finally, Sec. \ref{s:discussion} concludes the paper with some final remarks.
\section{Quantum estimation theory}  \label{s:QET}
Let us consider a family of quantum states $\varrho_\gamma$ parametrized by a parameter $\gamma$ that we want to estimate. If one performs a measurement described by a positive operator valued measure (POVM) $\{ \Pi_x \}$, the ultimate limit on the precision of any unbiased estimator for the parameter $\gamma$ is set by the Cram\'er-Rao bound \cite{CRBound}
\begin{align}
{\rm Var}(\gamma) &\geq 1/[M F(\gamma)]\:, \label{eq:QCRB}
\end{align}
where $M$ is the number of measurements performed,
$$
F(\gamma) = \int dx \: p(x|\lambda) \left( \partial_\gamma \log p(x|\gamma) \right)^2
$$ 
is the so-called (classical) Fisher information (FI), and $p(x|\gamma) = \Tr[\varrho_\gamma \Pi_x]$ denotes the conditional probability describing the whole measurement process. 
By optimizing over all the possible POVMs, one derives the Quantum Cram\'er-Rao bound (QCRB) \cite{QCRBound}
\begin{align}
{\rm Var}(\gamma) &\geq 1/[M F(\gamma)] \geq 1/[M H(\gamma)] \:, \label{eq:QCRB}
\end{align}
where  $H(\gamma) = \Tr[\varrho_\gamma L_\gamma^2]$ is the Quantum Fisher Information (QFI) and $L_\gamma$ is the symmetric logarithmic derivative (SLD) that is implicitly defined by the equation
$
2 \: \partial_\gamma \varrho_\gamma = L_\gamma \varrho_\gamma + \varrho_\gamma L_\gamma 
$.
As apparent from Eq. (\ref{eq:QCRB}), the QFI quantifies with how much precision one can estimate the parameter $\gamma$ independently from the specific measurement performed. Geometrically, the QFI corresponds to the Bures metric in the Hilbert space \cite{BuresFisher}: large values of the QFI correspond to large Bures distances between two quantum states $\varrho_\gamma$ and $\varrho_{\gamma + {\rm d}\gamma}$ obtained via an infinitesimal variation of the parameter $\gamma$. 
We also remark that for single-parameter estimation it is guaranteed that an optimal POVM saturating the QCRB always exists \cite{MatteoIJQI}.\\
\section{The dynamics}  \label{s:dynamics}
To fix ideas, we shall consider a single noisy continuous variable quantum degree of freedom subject to 
a positive definite harmonic Hamiltonian and to momentum diffusion, as would be the case 
for a trapped nanosphere undergoing heating via photon scattering, background gas collisions and 
blackbody radiation \cite{Oriol1,Rodenburg}. In order to simplify 
our treatment, we shall not include the typically smaller effects of position diffusion and friction \cite{Rodenburg}, 
which could be accounted for promptly within our formalism but would not add much conceptual insight. 
The additional stochastic term acting on the state vector according to the QMUPL model is equivalent to a momentum diffusion 
Lindblad superoperator entering the master equation for the quantum state $\varrho$. 
The same is approximately true for the centre of mass of motion of mesoscopic objects in the CSL model, 
since the position fluctuations are expected to be much smaller than the model localization length and one can perform a first order expansion of the 
superoperator \cite{nimmrichter}. Hence, the overall dynamics we shall consider is the following:
\begin{align}
\frac{{\rm d}\varrho}{{\rm d}t} &= \mathcal{L}\varrho 
= - i [\hat{H},\varrho] + \Gamma \: \mathcal{D}[\hat{x}]\varrho \; ,
\label{eq:ME}
\end{align}
where $[\hat{x},\hat{p}]=i$ ($\hbar=1$), 
$\hat{H}= \omega_m (\hat{x}^2 + \hat{p}^2)/2$ and $\mathcal{D}[\hat{O}]\varrho=O\varrho O^\dag - (O^\dag O \varrho + \varrho O^\dag O)/2$. The momentum diffusion rate is the sum of two contributions: $\Gamma= \Gamma_{\sf env} + \Gamma_{\sf fun}$, 
where $\Gamma_{\sf env}$ is due to environmental effect, while $\Gamma_{\sf fun}$ is fundamental. 
Our aim is analysing the estimation of $\Gamma_{\sf fun}$. 
Notice that such a parameter is equivalent to the only fundamental parameter of QMUPL and constrains the  
parameters of the CSL models through the formula
$\Gamma_{\sf fun} = \alpha \hbar \lambda_{\sf csl}/(m \omega_m r_c^2)$ \cite{nimmrichter}, 
where $m$ is the mass of the object, $\alpha$ is a factor that depends on its geometry, and the parameters $r_c$ and $\lambda_{\sf csl}$ characterise the model (the value for the intrinsic length scale is typically chosen at $r_c \approx 100$ nm, while bounds on the collapse rate are currently placed at $\lambda_c = 10^{-8 \pm 2} \textrm{s}^{-1}$ \cite{Adler}). 
Since we will consider only one mode, we will assume that after having trapped the nanosphere and cooled its motion down by sideband cooling, one will either turn off or detune the driving field in order to decouple the nanosphere motion and the cavity field. Hence, we will focus on the evolution of the mechanical oscillator alone. 
\subsection{Time-continuous monitoring} 
As already argued, in principle one can always counter the environmental decoherence by monitoring the environment. Here, we suppose to monitor the nanosphere position through the scattered light, obtaining a conditional dynamics described 
by the following conditional master equation \cite{Gardiner,WisemanMilburn}:
\begin{align}
{\rm d}\varrho = \mathcal{L}\varrho \: {\rm d}t + \sqrt{\eta \Gamma_{\sf env}} \mathcal{H}[\hat{x}] \varrho \: {\rm d}w \:,
\label{eq:SME}
\end{align}
where $\eta$ denotes the monitoring efficiency, $\mathcal{H}[O]\varrho = O \varrho + \varrho O^\dag - \Tr[\varrho ( O + O^\dag)]\varrho$ and ${\rm d}w$ represents a standard Wiener increment. As the Hamiltonian $\hat{H}$ is quadratic in the position and momentum operators, the evolution described by Eq. (\ref{eq:SME}) sends Gaussian states into Gaussian states, and thus we can fully describe it by looking at the evolution of first and second moments. Remarkably, due to a very specific property of quantum and classical conditional 
Gaussian statistics, one obtains \cite{WisemanDoherty,Diffusone} that, while the evolution of the first moments is, as expected, stochastic, {\em i.e.} depends on the results of the measurement performed on the environment, the covariance matrix evolves deterministically (the details of the Gaussian dynamics are explicitly given in the Appendix \ref{appB}).
Notice that the efficiency parameter $\eta\in[0,1]$ will allow us to describe realistic situations where the environmental degrees of freedom are only partially accessible (as would be the case for the imperfect collection of light scattered by a nanosphere). In principle there is no need to put an upper bound on $\eta$, and this is particularly the case for levitating nanospheres, that have been studied and proposed for the possibility of performing precise measurement via the direct monitoring of the trapping light scattered from the nanosphere itself \cite{Libbrecht}. On the other hand, in several experimental implementations as \cite{Rodenburg}, an additional ``environment'', completely under control and measurable, is added to the system, as a measurement device. This will also cause an extra diffusion rate $\Gamma_{\sf dev}$ and in this case the efficiency is upper bounded as $\eta \leq \Gamma_{\sf dev} / (\Gamma_{\sf dev} + \Gamma_{\sf env})$. Note also that setting $\eta=0$ obviously yields the original, unmonitored dynamics. As we show in Appendix \ref{appB}, in this case the dynamics does not admit a steady-state: this is due to the fact that, according to the unconditional master equation (\ref{eq:ME}), there is no damping acting on the mechanical oscillator.

For finite monitoring efficiency, thanks to the deterministic evolution of the second moments, one can prove that real-time linear feedback ({\em i.e.} real-time displacement in phase space depending on the measurement current) can be applied in order to obtain a steady state with zero first moments and thus remove the stochasticity of the evolution \cite{WisemanDoherty}. We remark here 
that this linear displacements will affect only the first moments evolution, while the evolution of the covariance matrix will still correspond to the one obtained via the stochastic master equation (\ref{eq:SME}).
This may not be the optimal strategy in terms of the estimation of $\Gamma_{\sf fun}$, but we will nonetheless 
consider this regime in what follows and set the first moments to zero, as it does allow for a deterministic steady state and will 
let us illustrate the advantage granted by the monitoring with a very compact, entirely analytical treatment. 
Also, in an experiment, a stable steady state is certainly much more desirable than a stochastically fluctuating 
one, on which one would have to perform an optimal discriminating measurement that would also fluctuate stochastically. 
Now, if all the other dynamical parameters are known, 
the steady state solutions of the equations for the covariance matrix reported in the Appendix \ref{appB}
yield a family of Gaussian states 
with zero first moments parametrized by the different 
values of $\Gamma_{\sf fun}$.
To assess the effectiveness of our strategy, in the following we will calculate the quantum and classical Fisher information for the parameter $\Gamma_{\sf fun}$. 
%As we present more in detail in the Appendix \ref{appA}, these two quantities give us the ultimate bounds on the ultimate precision we can obtain in the estimation of the parameter $\Gamma_{\sf fun}$.\\
%
\section{Fundamental diffusion estimation}   \label{s:results}
As detailed above, we want to assess the estimation of the parameter $\Gamma_{\sf fun}$, whose information is encoded in the Gaussian steady state obtained through the monitoring described by Eq. (\ref{eq:SME}). In particular, we will focus only on the steady state covariance matrix solution $\sigmaCM_{\sf ss}$ of the Riccati equation (\ref{eq:evolutionCM}), as we can assume that the first moments will be equal to zero. It should be remarked here that the linear driving needed to set the first moments to zero does depend on the parameter $\Gamma_{\sf fun}$ we need to estimate. And so will the final optimal 
quantum measurement to be performed on the steady state. However we can invoke, 
as customary in {\em local} quantum estimation problems, a multi-step adaptive protocol in order to solve this possible conundrum: 
one can apply the optimal protocol valid for an initial rough guess of the parameter $\Gamma_{\sf fun}$ 
(say, in our case, $\Gamma_{\sf fun}=0$), estimate the parameter through the measurement just performed, 
and then refine the operation and optimal measurement to be implemented given the latest estimate of the parameter. It has been shown in several cases that, after a few adaptive steps, one obtains 
an estimator giving the true value of the parameter and saturating the Cram\'er-Rao bound \cite{PhEstQubit,Berni2015}.

\begin{figure}[t!]
\begin{center}
\includegraphics[width=0.95\columnwidth]{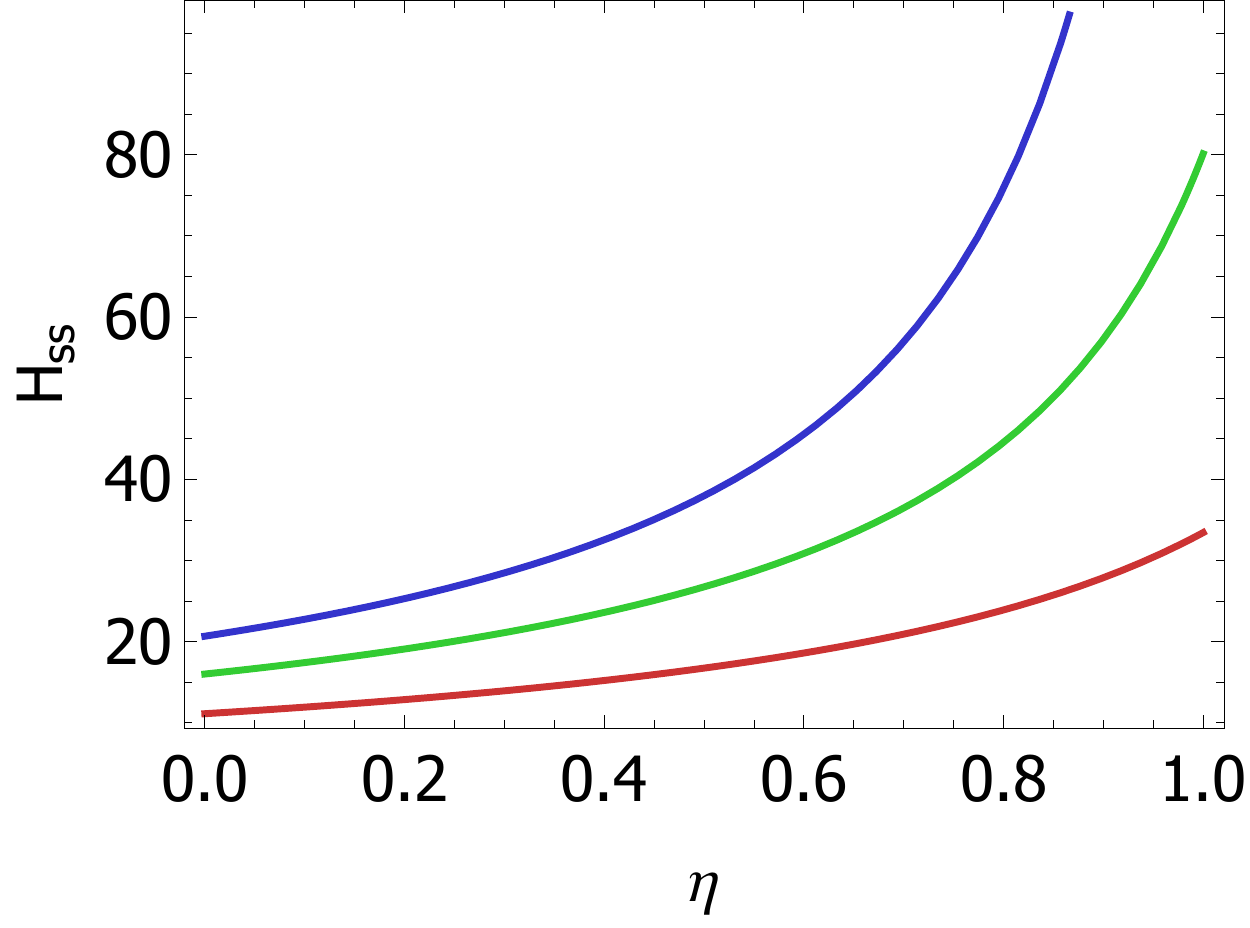} \hspace*{.2cm}
\end{center}
\caption{
QFI $H_{\sf ss}$ as a function of the monitoring efficiency $\eta$, for $\omega_m=1$, $\Gamma_{\sf env}=\omega_m/10$ and
and for different values of the estimated fundamental diffusion $\Gamma_{\sf fun}$. From top to bottom: $\Gamma_{\sf fun} = \{ \omega_m/100, \omega_m/40, \omega_m/20 \}$.
\label{f:QFI}}
%Right: Ratio between the classical FI for the POVM introduced in the main text (optimal for perfect monitoring and no collapses), and the QFI $H_{\sf ss}$ as a function of the monitoring efficiency $\eta$ and of the ratio between fundamental decoherence $\Gamma_{\sf fun}$ and environmental decoherence $\Gamma_{\sf env}$ (the environmental decoherence is kept fixed to $\Gamma_{\sf env}=\omega_m/10$).
%\label{f:optimalPOVM}}
\end{figure}
\begin{figure}[t!]
\begin{center}
\includegraphics[width=0.95\columnwidth]{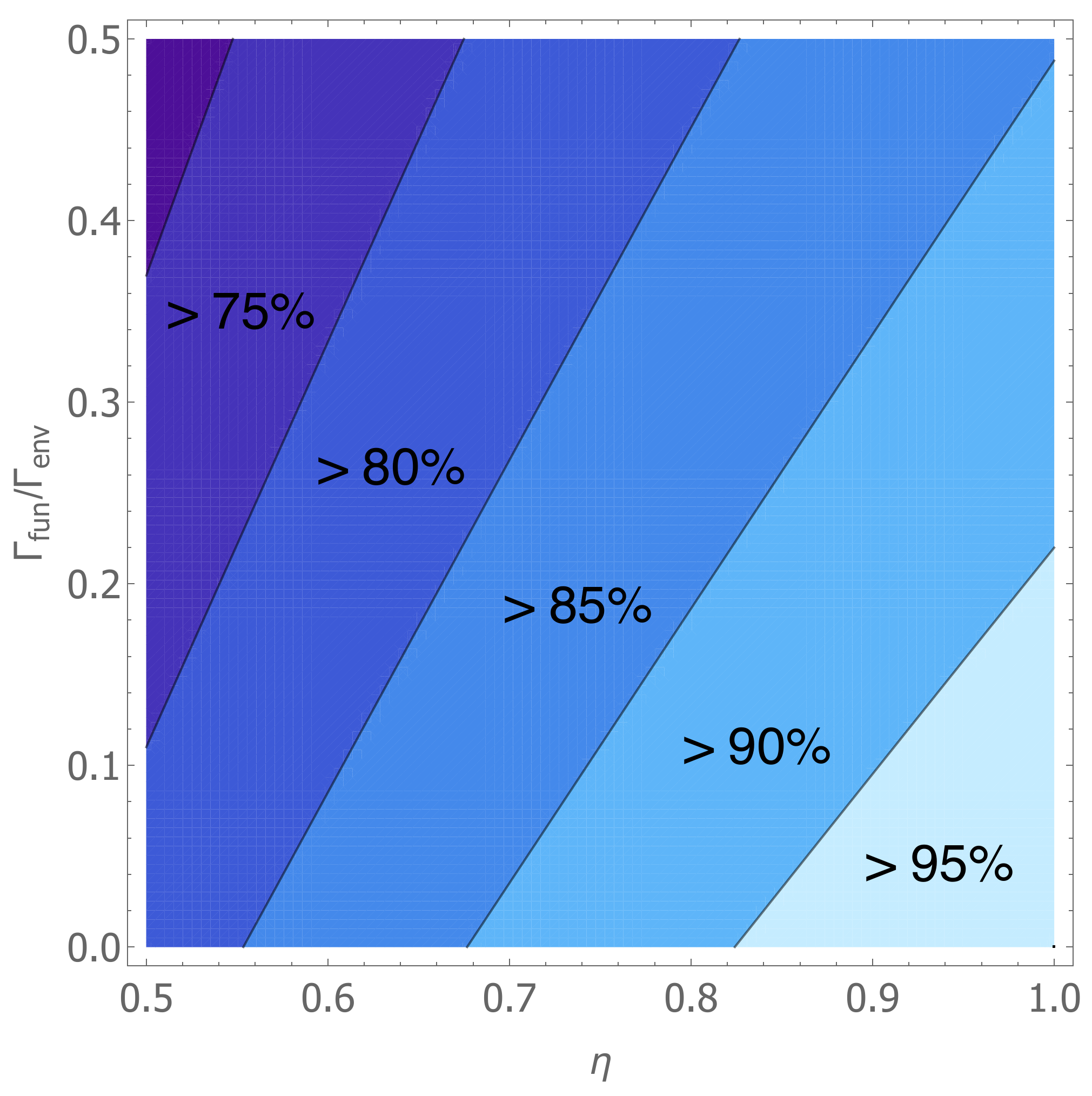}
\end{center}
\caption{
%QFI $H_{\sf ss}$ as a function of the monitoring efficiency $\eta$, for $\omega_m=1$, $\Gamma_{\sf env}=\omega_m/10$ and
%and for different values of the estimated fundamental diffusion $\Gamma_{\sf fun}$. From top to bottom: $\Gamma_{\sf fun} = \{ \omega_m/100, \omega_m/40, \omega_m/20 \}$.
%\label{f:QFI}
Ratio between the classical FI for the POVM introduced in the main text (optimal for perfect monitoring and no collapses), and the QFI $H_{\sf ss}$ as a function of the monitoring efficiency $\eta$ and of the ratio between fundamental decoherence $\Gamma_{\sf fun}$ and environmental decoherence $\Gamma_{\sf env}$ (the environmental decoherence is kept fixed to $\Gamma_{\sf env}=\omega_m/10$).
\label{f:optimalPOVM}}
\end{figure}

Under these assumptions (Gaussian steady state with zero first moments), the QFI can be evaluated analytically from the steady-state covariance matrix $\sigmaCM_{\sf ss}$, as described in \cite{AlexQFI,Pinel}.
The general analytical formulae for the steady state covariance matrix $\sigmaCM_{\sf ss}$ and the QFI $H_{\sf ss}$ are reported in the Appendix \ref{appB}. 
We start our analysis by discussing the general properties of the QFI by looking at Fig. \ref{f:QFI}. Firstly, it is apparent from the graph 
that the QFI increases monotonically with the detection efficiency $\eta$, providing one with a quantitative confirmation that countering the environmental decoherence through continuous measurements would help in the estimation of the collapse-induced diffusion. This proves also that our approach is useful in those cases where an additional environment has to be added as a measurement device \cite{Rodenburg}, and as a consequence $\eta$ is bounded because of these limitations of the experimental setup.
Also, the QFI decreases monotonically with $\Gamma_{\sf fun}$, implying that, in principle, 
smaller values of the parameter can be estimated more efficiently in terms of absolute error.
Next, it is instructive to consider the 
limiting case of perfect monitoring efficiency ($\eta =1$), 
where the QFI takes the compact analytical form
\begin{equation}
H_{\sf ss}(\eta=1) =
\frac{3 + \frac{4 \Gamma_{\sf env}}{\Gamma_{\sf fun}} - \frac{\omega_m}{\sqrt{\omega_m^2 +4 \Gamma_{\sf env} (\Gamma_{\sf env} + \Gamma_{\sf fun})}}}
{8 (\Gamma_{\sf env} + \Gamma_{\sf fun})(2 \Gamma_{\sf env} + \Gamma_{\sf fun})} \:.  \label{eta1}
%\\
%H_{\sf ss}(\Gamma_{\sf fun}=0) &= \frac{(3 + \eta)\sqrt{\omega_m^2 + 4 \Gamma_{\sf env}^2 \eta} - \omega_m (1-\eta)}{8 \Gamma_{\sf env}^2 (1-\eta^2)\sqrt{\omega_m^2 + 4 \Gamma_{\sf env}^2 \eta}} \:.
\end{equation}
One can easily check that in the limit of both perfect monitoring and zero fundamental decoherence ($\Gamma_{\sf fun}=0$), the QFI diverges. 
This can be intuitively understood by looking at how the steady state $\varrho_{\sf ss}$ changes in the Hilbert space by varying the parameter of interest: as we have already remarked, for perfect monitoring and zero fundamental decoherence, the steady state is pure; however, increasing the value of $\Gamma_{\sf fun}$ from zero introduces a diffusion that cannot be neutralized by monitoring the environment and, therefore, a mixed steady state (the same reasoning applies to the case of fixed $\Gamma_{\sf fun}=0$ and measurement efficiency $\eta$ decreasing from the maximum value). The abrupt change from pure to mixed states is responsible for the diverging QFI, 
and also yields insight as to the identification of the optimal quantum measurement saturating the quantum Cram\`er-Rao bound (QCRB), {\em i.e.} whose classical Fisher information is equal to the QFI. 
In point of fact, this argument singles out a dichotomic measurement corresponding to projecting either on the steady state itself $|\psi_{\sf ss}\rangle$ or on the rest of the Hilbert space, in order to be sensitive to the change from a pure to a mixed state. The corresponding POVM, described by the operators $\Pi_0 = |\psi_{\sf ss} \rangle \langle \psi_{\sf ss}|$ and $\Pi_1 = \mathbbm{1} - \Pi_0$, is indeed optimal, as 
it can be shown to achieve the quantum Cram{\'e}r-Rao bound. 
This POVM can be realized by first applying the symplectic operation that sends 
the vacuum state into the pure steady state $|\psi_{\sf ss}\rangle$
(as one can observe from the analytical solution for $\sigmaCM_{\sf ss}$, this operation involves some squeezing, 
that could be obtained by modulating the trap potential \cite{heinzen,TufoSoton}), 
and then by performing a vacuum projection (which, in optomechanics, could in principle be achieved by a
mapping of the mechanical state onto the light mode through red sideband driving, followed by a measurement 
of the latter with an avalanche photodiode, that distinguishes between zero and any positive number of photons).\\
The limits of perfect continuous monitoring of the environment and zero fundamental decoherence are clearly idealizations. However, the continuity of the QFI assures us that very high precision can be obtained in their neighbourhood, {\em i.e.} for high but not perfect efficiency $\eta\lesssim 1$ and for the interesting case of small CSL decoherence, when $\Gamma_{\sf fun}/\Gamma_{\sf env} \ll 1$. 
We have in fact also investigated the performance of the POVM just described also for different values of $\eta$ and $\Gamma_{\sf fun}$, where we know it is no longer optimal. As apparent from Fig. \ref{f:optimalPOVM}, the ratio between the classical FI and the QFI is still above $95\%$ for a reasonably large region of values of $\eta$ and $\Gamma_{\sf fun}$. This is particularly relevant since the optimal measurement, as is often the case, depends on the parameter to be estimated; given these results, one can indeed apply the optimal measurement for the case of zero fundamental decoherence, and still achieve a very high precision in the most interesting region of small, but not zero, values of $\Gamma_{\sf fun}$.
\subsection{Signal-to-noise ratio and efficient estimation} 
\noindent
More revealing than the quantum Fisher information itself is the associated signal to noise ratio
defined as the ratio between the estimated value and its standard deviation, {\em i.e.}
\begin{align}
S :=\frac{ \Gamma_{\sf fun}} { \sqrt{{\rm Var}(\Gamma_{\sf fun})}} \:.
\end{align}
A bound on the signal-to-noise ratio can be easily determined from Eqs.~(\ref{eq:QCRB}) and (\ref{eta1}) in the case of perfect monitoring, and reads
\begin{equation}
S_{\eta=1} \leq \sqrt{M} \frac{\Gamma_{\sf fun}}{\Gamma_{\sf env}} \sqrt{\frac{3+4 \frac{\Gamma_{\sf env}}{\Gamma_{\sf fun}}-\sqrt{\frac{1}{1+4\frac{\Gamma_{\sf env}^2}{\omega_{m}^2}+4\frac{\Gamma_{\sf fun}\Gamma_{\sf env}}{\omega_m^2}}}}
{8\left(1+\frac{\Gamma_{\sf fun}}{\Gamma_{\sf env}}\right)\left(2+\frac{\Gamma_{\sf fun}}{\Gamma_{\sf env}}\right)}}
\end{equation}
which, in the regime $\Gamma_{\sf fun}/\Gamma_{\sf env}\ll1$, goes like $\sqrt{M \Gamma_{\sf fun}/(4 \Gamma_{\sf env})}$
(let us remind the reader that $M$ is the number of estimation runs), 
and hence vanishes as $\Gamma_{\sf fun}$ vanishes.

\begin{figure}[t!]
\begin{center}
\includegraphics[width=0.95\columnwidth]{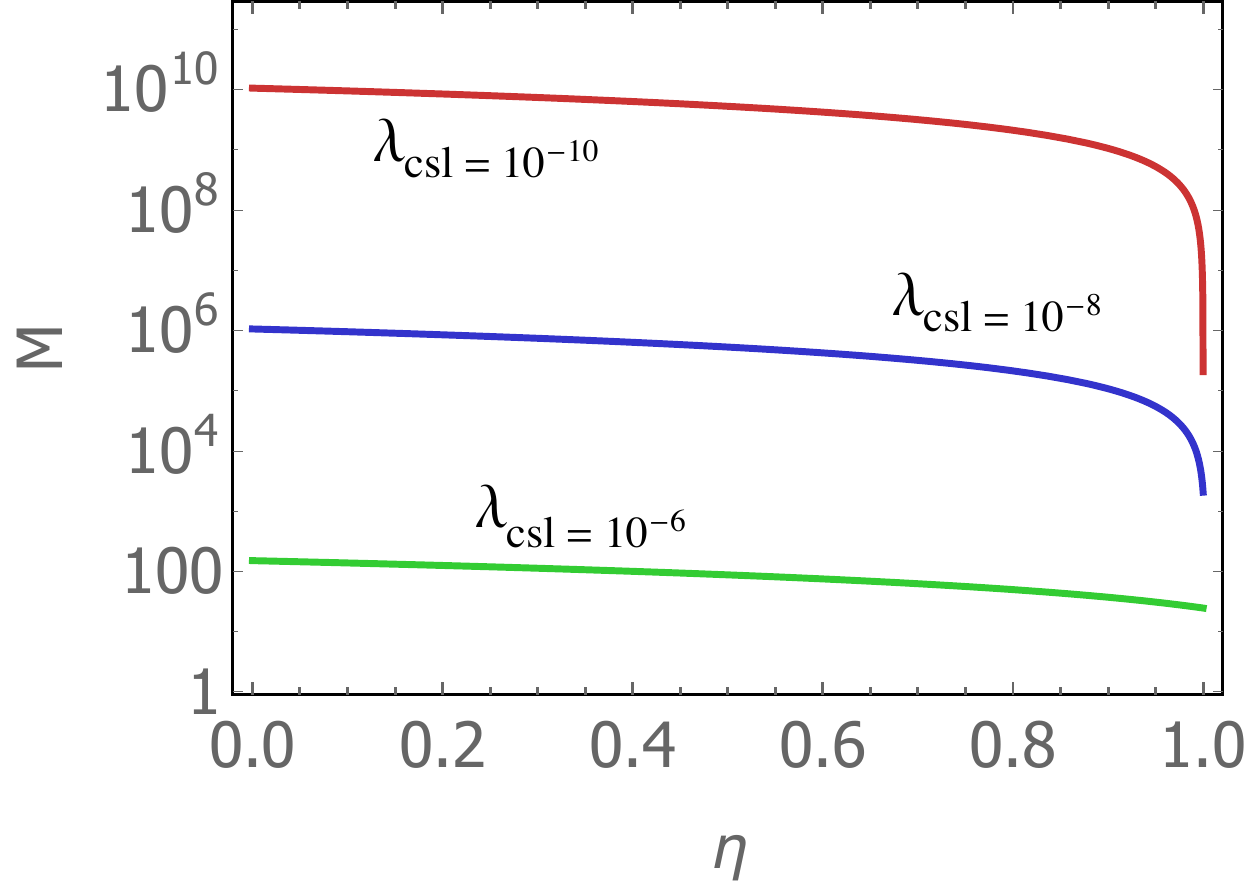}
\end{center}
\caption{
Measurements needed to get a signal-to-noise ratio larger than one as function of $\eta$ and for different values of $\lambda_{\sf csl}$ (other parameters are set considering a nanosphere of radius $r=r_c=100 \textrm{nm}$, with mechanical frequency $\omega_m/(2\pi)= 135$ kHz and with environmental diffusion $\Gamma_{\sf env}/(2\pi)=11$ kHz) . 
\label{f:SNR}}
%Right: Ratio between the QFI $H_t$ obtained for the state at time $t$ and the steady state QFI $H_{\sf ss}$ as a function of the evolution time $t$, for $\omega_m/(2\pi) = 135$ kHz, $\Gamma_{\sf env}/(2\pi)=11$ kHz and $\Gamma_{\sf fun}=10^{-5} \omega_m$. \label{f:finitetime}}
\end{figure}

As a further piece of analysis, one can consider the dependence of the ultimate signal to noise ratio $S^{\sf (csl)}$ 
on the mechanical frequency $\omega_m$
in the CSL model, where $\omega_m$ and the oscillator mass $m$ determine $\Gamma_{\sf fun}$. For perfect efficiency, this 
reads
\begin{align}
S_{\eta=1}^{\sf (csl)} &\leq  \sqrt{M}\beta \sqrt{ \frac{3+
4 \frac{ \omega_m \Gamma_{\sf env}}{\beta} - \frac{\omega_m}{\sqrt{\omega_m^2+4\Gamma_{\sf env}(\Gamma_{\sf env} +
 \beta/\omega_{m})}}} {8 (\beta + \Gamma_{\sf env} \omega_m) (\beta + 2 \Gamma_{\sf env} \omega_m)}}\, ,
\end{align}
with $\beta=\frac{\alpha \hbar \lambda_{\sf csl}}{m r_c^2 }$. This is a decreasing function of $\omega_{m}$, 
confirming that oscillators at lower frequencies (which, however, are more challenging to cool down to the quantum regime)
would prove advantageous in this context, as already indicated in proposals such as \cite{Goldwater15}, where an 
ion trap, rather than optical tweezers, was considered.\\
The most convincing evidence for the advantage granted by the continuous monitoring comes from considering the number of measurements
needed to achieve a signal to noise ratio of one as a function of the monitoring efficiency $\eta$. We considered
 the plausible scenario of continuous spontaneous localization with $\lambda_{\sf csl}=10^{-8 \pm 2}$, 
for a hypothetical nanosphere of radius $r=r_c=100 \textrm{nm}$, mechanical frequency $\omega_m/(2\pi)= 135$ kHz and subject environmental diffusion $\Gamma_{\sf env}/(2\pi)=11$ kHz, and we report the results in Fig.~\ref{f:SNR}. It can be seen that, for example fixing $\lambda_{\sf csl}=10^{-8}$,
the number of runs goes from around one million for the unmonitored case to around $3\times10^3$ for perfect monitoring.
Hence, one concludes that environmental monitoring would help in designing experiments able to improve the existing bounds 
on $\Gamma_{\sf fun}$.
\begin{figure}[t!]
\begin{center}
\includegraphics[width=0.95\columnwidth]{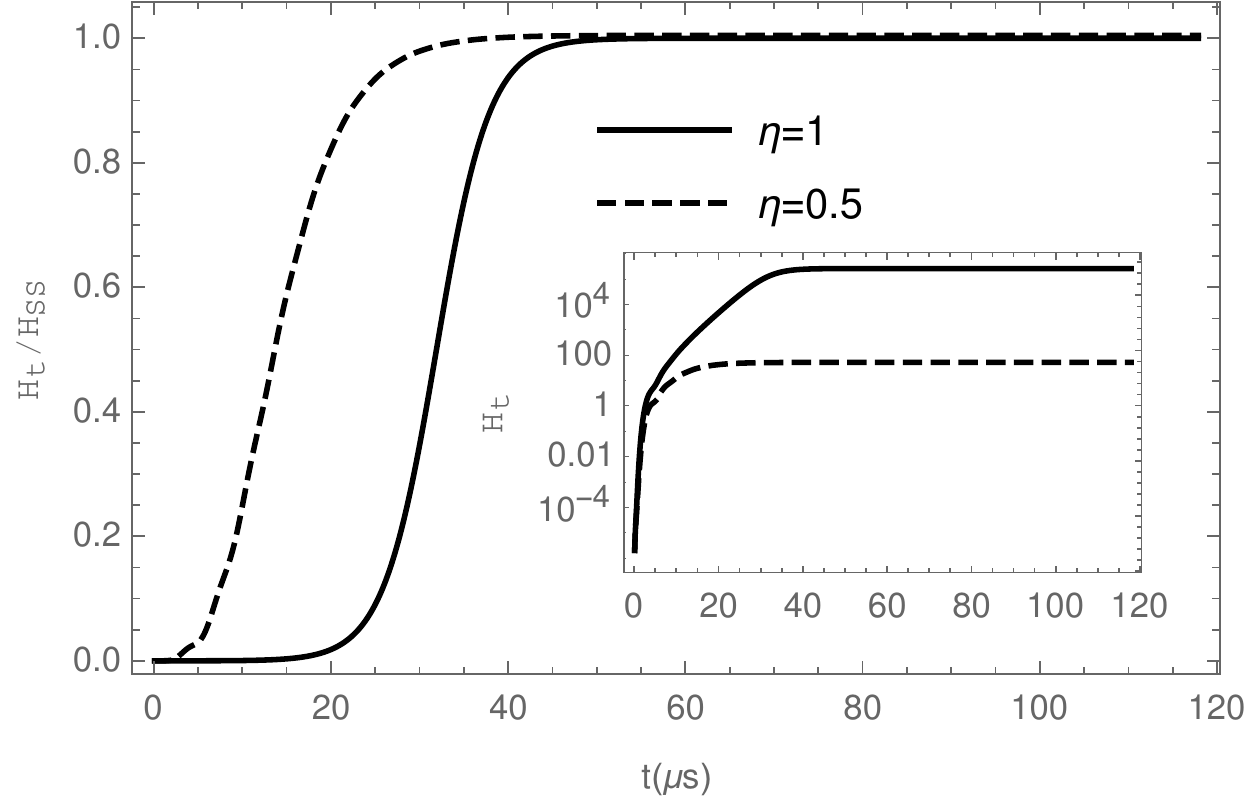}
\end{center}
\caption{
%Left: Measurements needed to get a signal-to-noise ratio larger than one as function of $\eta$ and for $\lambda_{\sf csl}=10^{-8}$ (other parameters are set considering a nanosphere of radius $r=r_c=100 \textrm{nm}$, with mechanical frequency $\omega_m/(2\pi)= 135$ kHz and with environmental diffusion $\Gamma_{\sf env}/(2\pi)=11$ kZhz) . 
%\label{f:SNR}
Ratio between the QFI $H_t$ obtained for the state at time $t$ and the steady state QFI $H_{\sf ss}$ as a function of the evolution time $t$, for $\omega_m/(2\pi) = 135$ kHz, $\Gamma_{\sf env}/(2\pi)=11$ kHz and $\Gamma_{\sf fun}=10^{-5} \omega_m$, and for an initial thermal state with $n_{\sf th}=100$ average phonons. Inset: QFI $H_t$, in logarithmic scale, as a function of time (same parameters values and initial state of the main plot). \label{f:finitetime}}
\end{figure} 
\subsection{Finite time analysis} 
All the results reported above 
have been derived considering the mechanical oscillator steady state. 
In order to validate such an analysis, we investigate the transient dynamics of the QFI and in particular its ratio with the QFI obtained at steady state. The results are plotted for experimentally reasonable values of the parameters in Fig.~\ref{f:finitetime}. 
We focus in particular on small values of the fundamental decoherence parameter $\Gamma_{\sf fun}$, but we have numerical evidence that similar results are obtained for larger values. As can be seen from the graph, the ratio goes indeed to one in a relatively small time (with our parameters, around $30 \mu\textrm{s}$ for efficiency $\eta=0.5$ and around $50 \mu\textrm{s}$ for perfect measurement, $\eta=1$), {\em i.e.} the steady-state precision on the estimation of the parameter $\Gamma_{\sf fun}$ can be safely obtained at finite time. It should be also noticed from the inset, that even at finite time, the QFI for perfect efficiency is always larger than the one corresponding to finite efficiency ($\eta=0.5$ in our plot).
\section{Discussion}  \label{s:discussion}
It should be noted that our estimation analysis assumed 
perfect knowledge of all the dynamical parameters other than our target $\Gamma_{\sf fun}$. 
This may be particularly delicate, especially in regard to the environmental diffusion $\Gamma_{\sf env}$, 
which would have to be inferred from other parameters through theoretical considerations \cite{Chang2010}. 
Quantum estimation theory could however be adapated to allow for uncertainties in 
dynamical parameters other than the estimated one (see e.g. \cite{Mihai2014,AnimeshMultiMetro}).
In the optomechanical paradigm, one could also take into account the coupled light field: 
the extension of the present study to the full, two-mode optomechanical system, 
and the identification of associated optimal global detection strategies, will be 
an interesting development of this line of inquiry. 
Moreover one can also investigate how to exploit the information obtained through the time-continuous monitoring in order to improve the estimation of the parameters of interest, as, for example described in \cite{Mankei}, where the time-continuous estimation of a classical stochastic process coupled to a dynamical system is studied in detail.
Regardless of such issues, which are common to all investigations into fundamental decoherence, it is crucial to remark that 
the schemes we described have the power to {\em falsify} wave function collapse theories, in the sense that they can rule out regions 
in the noise parameters space by setting upper bounds to the diffusion rates, which hold even in the presence of unknown additional noise.
In this regard, our study unambiguously highlights the substantial advantage that monitoring environmental decoherence would grant.
Notice also that the practical feasibility of such monitoring is certainly within current experimental capabilities, since 
it has already been demonstrated to other aims in several set-ups where our formalism applies \cite{KalmanAspelmeyer,Lia2016}. 
Our suggestion is hence timely and has potential for immediate practical impact in the effort to falsify collapse theories.
In our scheme monitoring would be, in a sense, an active way of ``putting aside the impediments of matter'' that hinder 
the detection of fundamental effects, much in the same fashion as friction was standing 
in the way of Galileo's analysis of free fall motion \cite{galileo}.
\section{Acknowledgments} 
\noindent
We thank P. Barker and D. Goldwater for discussions.
MGG and AS acknowledge support
from EPSRC through grant EP/K026267/1. MGG acknowledges
support from the Marie Sk\l odowska-Curie Action H2020-MSCA-IF-2015 (project ConAQuMe). OSD acknowledge support from grant 2015/12747-9, S\~{a}o Paulo Research Foundation (FAPESP).\\
{\em Note added}: after the completion of this work, we became aware of a related analysis where the QFI is employed to assess non-interferometric tests of wave-function collapse models (see S. McMillen et al., arXiv:1606.00070).
\appendix
\section{Analytical solution of the Gaussian dynamics} \label{appB}
Here we will provide the formulas describing the time-evolution of the mechanical oscillator under time-continuous monitoring as prescribed by the stochastic master equation (\ref{eq:SME}). As we mentioned above, the whole dynamics preserve the Gaussian character of the quantum state and thus can be fully described in terms of the first moments vector $\langle {\bf \hat{r}}\rangle$ and of the covariance matrix $\sigmaCM$ of the quantum state $\varrho$, defined in components as $\langle \hat{r}_j\rangle={\rm Tr}\left[\hat{r}_j \varrho\right]$ and $\sigma_{jk}={\rm Tr}\left[\{\hat{r}_j - \langle \hat{r}_j \rangle ,\hat{r}_k - \langle \hat{r}_k \rangle\}\varrho\right]$ for the operator vector $\hat{\bf r}=(\hat{x},\hat{p})^{\sf T}$. 
In formulae one obtains \cite{Diffusone,WisemanDoherty}:
\begin{align}
d\langle {\bf \hat{r}}\rangle &= A \langle {\bf \hat{r}}\rangle  dt - \sigmaCM B d{\bf w} \:,  \\
\frac{d \sigmaCM}{dt} &= A \sigmaCM + \sigmaCM A^{\sf T} + Q -\sigmaCM B B^{\sf T} \sigmaCM  \:,  \label{eq:evolutionCM}
\end{align}
where $d{\bf w}$ is a vector of Wiener increments such that $\{ d{\bf w}, d{\bf w}^{\sf T} \} = \mathbbm{1} dt$ and the matrices read
\begin{align}
A &=
\left(
\begin{array}{c c} 
0 & \omega_m \\ 
-\omega_m & 0
\end{array}
\right), \\
Q &=
\left(
\begin{array}{c c} 
0 & 0 \\
0 & 2 (\Gamma_{\sf env} + \Gamma_{\sf fun}) \\
\end{array}
\right) , \\
B &=
\left(
\begin{array}{c c} 
0 & \sqrt{2 \eta \Gamma_{\sf env}} \\
0 & 0  
\end{array}
\right) .
\end{align}
The existence of a steady-state for a continuously monitored quantum systems has been discussed in \cite{WisemanDoherty}. It is shown that Eq. (\ref{eq:evolutionCM}) has a stabilizing solution if and only if the pair of matrices $(B,{A})$ is {\em detectable}, {\em i.e.}
\begin{align}
B {\bf x}_\lambda \neq 0 \:\:\:\: \forall \: {\bf x}_\lambda : {A}{\bf x}_\lambda = \lambda {\bf x}_\lambda \:\: \textrm{with} \: {\rm Re}[\lambda] \geq 0 \:, 
\end{align}
that is whenever information on the degrees of freedom that are not strictly stable under the drift matrix ${A}$, is obtained in the measurement output $B \langle \hat{\bf r}\rangle $. We find that, in our system, this condition is met for every non-zero efficiency $0<\eta\leq 1$. It is on the other hand known that, for no monitoring ($\eta=0$), the system is not stable as no damping terms are present in the drift matrix $A$.\\
The steady state covariance matrix can be derived analytically as
\begin{align}
\sigmaCM_{\sf ss} &=
\left(
\begin{array}{c c} 
\frac{\sqrt{\omega_m \left( \Upsilon - \omega_m \right)}}{\sqrt{2} \eta \Gamma_{\sf env}} & \frac{\Upsilon - \omega_m}{2 \eta \Gamma_{\sf env}} \\
 \frac{\Upsilon - \omega_m}{2 \eta \Gamma_{\sf env}} & 
\frac{\Upsilon \sqrt{\left( \Upsilon - \omega_m \right)}}{\sqrt{2 \omega_m} \eta \Gamma_{\sf env}} 
\end{array}
\right), \label{eq:sscov}
\end{align}
where 
\begin{align}
\Upsilon = \sqrt{\omega_m^2 + 4 \eta \Gamma_{\sf env} (\Gamma_{\sf env} + \Gamma_{\sf fun})} \:.
\end{align}
As a further proof that this covariance matrix can be obtained at steady-state via continuous measurements and feedback, one can easily check that, for $0<\eta \leq 1$, it satisfies the necessary and sufficient condition derived in \cite{WisemanDoherty}: $A \sigmaCM_{\sf ss} + \sigmaCM_{\sf ss} + Q \geq 0$.\\
Notice that, typically, the steady state above is a squeezed state, in the sense that its smallest eigenvalue will be smaller than one. Obtaining the decomposition, in terms of diagonal single-mode squeezers and orthogonal phase shifters, of the symplectic operation that relates the vacuum state to this steady state  is a straightforward task, that just requires one to diagonalise 
the matrix $\sigmaCM_{\sf ss}$.\\

The corresponding QFI for the estimation of the parameter $\Gamma_{\sf fun}$ can be easily evaluated by using the formula \cite{Pinel}
\begin{align}
H_{\sf ss} = \frac12 \frac{\hbox{Tr}[(\sigmaCM_{\sf ss}^{-1}\sigmaCM_{\sf ss}^\prime)^2]}{1+\mu_{\sf ss}^2}
+ 2 \frac{(\mu_{\sf ss}^{\prime})^2}{1-\mu_{\sf ss}^4} \:, \label{eq:GaussQFI}
\end{align}
where $\mu_{\sf ss}=\Tr[\varrho_{\sf ss}^2] = 1/\sqrt{\det[\sigmaCM_{\sf ss}]}$ represents the purity of the state, and primed quantities corresponds to derivative with respect to the parameter $\Gamma_{\sf fun}$. By applying it to the steady-state covariance matrix in Eq. (\ref{eq:sscov}), one obtains
\begin{align}
H_{\sf ss} = \frac{
\Gamma_{\sf env} \left[ (1-\eta) \omega_m - (3+\eta) \Upsilon \right] + \Gamma_{\sf fun} \left( \omega_m - 3 \Upsilon\right)}
{8\Upsilon (\Gamma_{\sf env} + \Gamma_{\sf fun}) \left[ \eta^2 \Gamma_{\sf env}^2 - (\Gamma_{\sf env} + \Gamma_{\sf fun})^2 \right] }.
\end{align}
%
%
%
%%%%%%%%%%%%%%%%%%%%%%%%%%%%%

%%....
%

\begin{thebibliography}{99}
\bibitem{burgnote}Let us mention here that another, possibly complementary, strategy to distinguish between 
fundamental and environmental quantum noise could be offered 
by dynamical decoupling techniques \cite{burgarth1,burgarth2}.
\end{thebibliography}
\end{document}